# Improved Lower Bounds for Constant GC-Content DNA Codes

Yeow Meng Chee and San Ling

*Abstract*—The design of large libraries of oligonucleotides having constant GC-content and satisfying Hamming distance constraints between oligonucleotides and their Watson-Crick complements is important in reducing hybridization errors in DNA computing, DNA microarray technologies, and molecular bar coding. Various techniques have been studied for the construction of such oligonucleotide libraries, ranging from algorithmic constructions via stochastic local search to theoretical constructions via coding theory. A new stochastic local search method is introduced, which yields improvements for more than one third of the benchmark lower bounds of Gaborit and King (2005) for $n$-mer oligonucleotide libraries when $n \leq 14$. Several optimal libraries are also found by computing maximum cliques on certain graphs.

## I. INTRODUCTION

Oligonucleotides (short single-stranded DNA) made by chemical synthesis are important structures for information storage in DNA computing [1], [2], as probes in DNA microarray technologies [3], [4], and as tags in molecular bar coding [5]–[7]. The critical property of DNA in these applications is the tendency of oligonucleotides to specifically hybridize to their Watson–Crick complements and form a stable duplex [8].

Unfortunately, nonspecific hybridizations can also occur between oligonucleotides used in a self-assembly step, in a polymerase chain reaction, or in an extraction operation. The probability of such hybridization errors is related to the combinatorial as well as the thermodynamic properties of the oligonucleotides. Among the basic constraints that must be fulfilled in order to reduce the probability of erroneous hybridizations for a library of oligonucleotides, the following are of particular importance:

1) two oligonucleotides in the library must be dissimilar;
2) an oligonucleotide in the library must be dissimilar to the (Watson–Crick) complement of another oligonucleotide in the library;
3) every oligonucleotide in the library has similar melting temperature;
4) an oligonucleotide must not fold back onto itself in a manner that renders it chemically inactive.

The measure of similarity between oligonucleotides depends on the hybridization model adopted. On two extremes of the spectrum, we have the following.

Manuscript received October 19, 2006; revised September 2, 2007. This work was supported in part by the Singapore Ministry of Education under Research Grant T206B2204.

Y. M. Chee is with the Interactive Digital Media R&D Program Office, Media Development Authority, Singapore 179369, Republic of Singapore, the Division of Mathematical Sciences, School of Physical and Mathematical Sciences, Nanyang Technological University, Singapore 637616, Republic of Singapore, and with the Department of Computer Science, School of Computing, National University of Singapore, Singapore 117590, Republic of Singapore (e-mail: ymchee@alumni.uwaterloo.ca).

S. Ling is with the Division of Mathematical Sciences, School of Physical and Mathematical Sciences, Nanyang Technological University, Singapore 637616, Republic of Singapore (e-mail: lingsan@ntu.edu.sg).

- *Hamming distance model* [9]–[11]. The sugar-phosphate backbone of oligonucleotides is nonelastic, and an oligonucleotide can only hybridize to its Watson–Crick complement.
- *Levenshteǐn distance model* [12], [13]. The sugar-phosphate backbone of oligonucleotides is completely elastic, and an oligonucleotide $\sigma$ can hybridize to any oligonucleotide containing the Watson–Crick complement of $\sigma$ as a subsequence.

In actual fact, the sugar-phosphate backbone of oligonucleotides shows some limited elasticity, and the stability of a hybridized duplex is determined by the nearest neighbor interaction energies and stacking energies of the hybridized bases [14], which are difficult to model accurately with purely combinatorial constraints. Hybridization models based on thermodynamical properties of oligonucleotides have been proposed as better approximations [15]. Other measures of similarity between oligonucleotides have also been considered [16]. Recently, Chen *et al.* [17] addressed the problems of predicting hybridization properties of long oligonucleotides. In short, the problem of what properties oligonucleotides have to possess in order to exhibit very specific hybridization behavior is not well understood, except those of short lengths.

The model we adopt in this correspondence is the Hamming distance model. It should be noted that the constraints and the hybridization model we consider do not address certain issues related to hybridization which may be important in practical applications, for example insensitivity to frame-shifts, the avoidance of secondary structures, and the use of a more accurate model of melting temperature [18]–[20]. Our model also does not consider DNA folding, which is one of the most important properties one has to test in the process of probe selection. However, for the sequence lengths that we consider in this correspondence, folding is not expected to be severe (not too many oligonucleotides of up to 8-mers fold, and even if they do, the folds are usually not very stable) [21].

For the purpose of efficiency in the applications mentioned above, it is desirable that for a given $n$, we have as large a library of $n$-mer oligonucleotides as possible that satisfies constraints 1) to 3) above. This is the *oligonucleotide (or DNA) sequence design problem* [9], [22]–[24]. Many approaches have been considered for this problem. These include template-based constructions [11], [24]–[26], stochastic local search [27]–[31], lexicographic search [9], and coding theoretic constructions [10]. A survey of the best lower bounds for the sizes of oligonucleotide libraries has been undertaken by Gaborit and King [10].

The purpose of this correspondence is to introduce a new stochastic local search method for the oligonucleotide sequence design problem. This search method has been implemented and yielded many record-breaking oligonucleotide libraries. Several optimal oligonucleotide libraries were also obtained via an exhaustive search algorithm based on computing maximum cliques on graphs.

## II. DEFINITIONS AND NOTATIONS

We model oligonucleotides as sequences over the alphabet $\Sigma = \{A, C, G, T\}$. If $\sigma \in \Sigma^n$, the element in position $i$ of the sequence $\sigma$ is denoted $\sigma_i$. The *Hamming distance* between two sequences $\sigma, \tau \in \Sigma^n$, denoted $d_H(\sigma, \tau)$, is the number of positions where $\sigma$ and $\tau$ differ, that is

$$d_H(\sigma, \tau) = |\{1 \leq i \leq n : \sigma_i \neq \tau_i\}|.$$

The (Watson–Crick) *complement* of a sequence $\sigma = \sigma_1 \ldots \sigma_n \in \Sigma^n$ is the sequence $\bar{\sigma} = \bar{\sigma}_n \ldots \bar{\sigma}_1 \in \Sigma^n$, where

$$\bar{A} = T, \bar{C} = G, \bar{G} = C, \bar{T} = A.$$

The GC-*content* of a sequence $\sigma \in \Sigma^n$, denoted $GC(\sigma)$, is the number of occurences of G and C in $\sigma$:

$$GC(\sigma) = |\{1 \leq i \leq n : \sigma_i \in \{\mathsf{C}, \mathsf{G}\}\}|.$$

Henceforth, lower case Greek letters are used to denote oligonucleotides, and if not otherwise stated, they are assumed to belong to a generic set $\mathcal{L}$.

A library of $n$-mer oligonucleotides $\mathcal{L} \subseteq \Sigma^n$ satisfying all the constraints.
1) *Hamming distance constraint*: $d_H(\sigma, \tau) \geq d$ for all $\sigma, \tau \in \mathcal{L}$, $\sigma \neq \tau$;
2) *Complementary distance constraint*: $d_H(\sigma, \bar{\tau}) \geq d$ for all $\sigma, \tau \in \mathcal{L}$;
3) *Constant GC-content constraint*: $GC(\sigma) = w$ for all $\sigma \in \mathcal{L}$;

is called an $(n, d, w)$-*DNA code*. Note that the second constraint has to hold also for $\tau = \sigma$. If $\mathcal{L} \subseteq \Sigma^n$ satisfies only the Hamming distance and the constant GC-content constraints, we call $\mathcal{L}$ a *weak* $(n, d, w)$-*DNA code*, Following King [9], we denote the maximum size of an $(n, d, w)$-DNA code by $A_4^{GC,RC}(n, d, w)$, and the maximum size of a weak $(n, d, w)$-DNA code by $A_4^{GC}(n, d, w)$. A (weak) $(n, d, w)$-DNA code containing $A_4^{GC,RC}(n, d, w)$ ($A_4^{GC}(n, d, w)$) sequences is said to be *optimal*. The following halving bound is known [9], [23].

*Lemma 1 (Marathe et al. King):* For $0 < d \leq n$ and $0 \leq w \leq n$

$$A_4^{GC,RC}(n, d, w) \leq \frac{1}{2} A_4^{GC}(n, d, w).$$

## III. THE DNA CODE DESIGN ALGORITHM

Stochastic local search algorithms for determining $(n, d, w)$-DNA codes of size $A$ typically adopt the following framework. We begin with a subset $\mathcal{L} \subseteq \Sigma^n$ and we iteratively modify $\mathcal{L}$ until we obtain an $(n, d, w)$-DNA code of size $A$. A modification step comprises in moving $\mathcal{L}$ to a random *neighbor* $\mathcal{L}'$, with an *acceptance probability* determined by its *proximity* to being an $(n, d, w)$-DNA code of size $A$. The art of designing stochastic local search algorithms for DNA codes lies in the specification of
1) a good initialization procedure;
2) $N(\mathcal{L})$, the *neighborhood* of $\mathcal{L}$;
3) $\mathrm{cost}(\mathcal{L})$, the *measure of proximity* of $\mathcal{L}$ to a solution;
4) $f$, the *acceptance probability function*; and
5) a reasonably efficient stopping criterion.

The best performing stochastic search algorithm for DNA codes currently is that of Tulpan *et al.* [30], [30]. In their algorithm, the starting configuration $\mathcal{L}$ is a random set of $A$ elements from $\Sigma^n$, each having GC-content $w$. The neighborhood $N(\mathcal{L})$ is defined to contain those subsets of $\Sigma^n$ obtained by "mutating" two sequences $\sigma, \tau \in \mathcal{L}$ that violate at least one of the Hamming distance or complementary distance constraints, hopefully to two sequences that violate fewer number of contraints. Several mutation strategies were considered by Tulpan *et al.* The proximity of $\mathcal{L} \subseteq \Sigma^n$ to an $(n, d, w)$-DNA code, $\mathrm{cost}(\mathcal{L})$, is the number of times a Hamming distance or complementary distance constraint is violated. A neighbor $\mathcal{L}' \in N(\mathcal{L})$ of $\mathcal{L}$ is always accepted if $\mathrm{cost}(\mathcal{L}') \leq \mathrm{cost}(\mathcal{L})$, and is accepted with probability $f(\mathrm{cost}(\mathcal{L}'))$ if it has a higher cost, to allow escape from local optima.

Our approach is orthogonal to that of a Gilbert–Varshamov-like construction, where the entire space $\Sigma^n$ is taken as the initial set, and conflicting DNA oligonucleotides are repeatedly removed until a set that contains nonconflicting DNA oligonucleotides is obtained. However, for large $n$ (say $n \geq 20$), this approach becomes computationally infeasible since we may not be able to generate the whole set of DNA oligonucleotides in reasonable time. Our approach starts with a small DNA code and progressively moves it toward a DNA code of target size (while maintaining full Hamming distance and complementary distance constraint satisfaction at all times). More specifically,
1) we start initially with $\mathcal{L}$ being the empty set;
2) for any $\mathcal{L} \subseteq \Sigma^n$, $N(\mathcal{L})$ is the set of all $\mathcal{L}'$ that is obtained from $\mathcal{L}$ by adding a new sequence $\sigma$ of GC-content $w$ which satisfies $d_H(\sigma, \bar{\sigma}) \geq d$ and removing all those $\tau$ from $\mathcal{L}$ that violates at least one of $d_H(\sigma, \tau) \geq d$ or $d_H(\sigma, \bar{\tau}) \geq d$.

The proximity, $\mathrm{cost}(\mathcal{L})$, is simply the number of sequences $\tau$ removed. The acceptance probability function we adopt has the form

$$f(x) = \begin{cases} 1, & \text{if } x \in \{0, 1\} \\ \alpha \exp(-x/\beta), & \text{if } x \in \{2, 3\} \\ 0, & \text{if } x \geq 4 \end{cases}$$

where $x = \mathrm{cost}(\mathcal{L})$, and $\alpha, \beta$ are constants. The function $f$ is designed so that a move to a solution that is at least as large as the current solution is always accepted, while a move to a solution that has size three or more smaller than the current solution is always rejected. The reason for this is that as the solution moves closer to an optimum, it is observed that it can take a long time to move back to a solution of equal size if we were to accept such a "drastic" downhill move. We also observed that $6 \cdot 10^{-5} \leq \alpha \leq 7 \cdot 10^{-5}$ and $1.4 \leq \beta \leq 1.5$ work quite well, although no comprehensive empirical studies were carried out to determine if these were the best settings for $\alpha$ and $\beta$. More rigorous analysis on the choices for $\alpha$ and $\beta$ can be conducted by studying the algorithm's *solution quality*, *run-length*, and *run-time* distributions. We refer the reader to [32] for details. The algorithm is terminated if there is no improvement to the size of the best $(n, d, w)$-DNA code obtained thus far, after a specified number of iterations $M$. A more detailed description of our algorithm is given in Fig. 1.

We note that our stochastic local search algorithm, with the specified acceptance probability function, is essentially the Metropolis algorithm [33]. It can also be considered a form of the simulated annealing algorithm [34] without a cooling schedule. The reader is referred to [32] for a systematic and unified treatment of stochastic local search algorithms.

```
Input: n, d, w, A, M
L ← ∅;
bestcode ← ∅;
bestsize ← 0;
iterations ← 0;
while (|L| ≠ A and iterations ≤ M) {
    σ ← random element of {σ ∈ Σⁿ : GC(σ) = w and d_H(σ, σ̄) ≥ d};
    T ← {τ ∈ L : d_H(σ, τ) < d or d_H(σ, τ̄) < d};
    cost(L) ← |T|;
    with probability f(cost(L)) {
        L ← (L \ T) ∪ {σ};
        if (|L| > bestsize) {
            bestcode ← L;
            bestsize ← |L|;
            iterations ← 0;
        }
    }
    iterations ← iterations + 1;
}
Output: (n, d, w)-DNA code bestcode L of size bestsize
```

Fig. 1. Stochastic local search algorithm for DNA codes.

## IV. OPTIMAL CODES AND MAXIMUM CLIQUES

To evaluate and improve stochastic search algorithms, it is important to have knowledge of optima for various parameters. In this section, we determine some optimal $(n, d, w)$-DNA codes, and hence the value of $A_4^{GC,RC}(n, d, w)$, computationally. We outline our approach below.

For given $n$, $d$, and $w$, we construct a graph $\Gamma_{n,d,w}^{GC,RC}$ as follows. The vertex set $V(\Gamma_{n,d,w}^{GC,RC})$ is the set of all sequences $\sigma \in \Sigma^n$ such that $GC(\sigma) = w$ and $d_H(\sigma, \bar{\sigma}) \geq d$. A pair $\{\sigma, \tau\}$ appears in the edge set $E(\Gamma_{n,d,w}^{GC,RC})$ if and only if $d_H(\sigma, \tau) \geq d$ and $d_H(\sigma, \bar{\tau}) \geq d$. It is easy to see that $\mathcal{L} \subseteq \Sigma^n$ is an $(n, d, w)$-DNA code if and only if $\mathcal{L}$ is a clique of $\Gamma_{n,d,w}^{GC,RC}$. Hence, an optimal $(n, d, w)$-DNA code corresponds to a maximum clique of $\Gamma_{n,d,w}^{GC,RC}$.

TABLE I
STATISTICS OF SOME $\Gamma_{n,d,w}^{GC,RC}$ AND $\Gamma_{n,d,w}^{GC}$

| $n$ | $d$ | $w$ | $|V(\Gamma_{n,d,w}^{GC,RC})|$ | $|E(\Gamma_{n,d,w}^{GC,RC})|$ | density | size of max clique | number of distinct max cliques |
|---|---|---|---|---|---|---|---|
| 5 | 3 | 2 | 304 | 34,848 | 0.75664 | 15 | 8,388,608 |
| 5 | 4 | 2 | 208 | 6,208 | 0.28837 | 3 | 16,384 |
| 6 | 4 | 3 | 864 | 223,176 | 0.59862 | 16 | 58,720,256 |
| 7 | 5 | 3 | 3,904 | 3,945,728 | 0.51790 | 11 | 446,693,376 |
| 7 | 6 | 3 | 2,224 | 241,664 | 0.09776 | 2 | 241,664 |
| $n$ | $d$ | $w$ | $|V(\Gamma_{n,d,w}^{GC})|$ | $|E(\Gamma_{n,d,w}^{GC})|$ | density | size of max clique | number of distinct max cliques |
| 5 | 3 | 2 | 320 | 44,800 | 0.87774 | 30 | 12,288 |
| 6 | 5 | 3 | 1,280 | 437,120 | 0.53401 | 8 | 248,709,120 |

TABLE II
LOWER BOUNDS FOR $A_4^{GC,RC}(n,d,\lfloor n/2 \rfloor)$

| $n \setminus d$ | 3 | 4 | 5 | 6 | 7 | 8 | 9 | 10 | 11 | 12 | 13 | 14 |
|---|---|---|---|---|---|---|---|---|---|---|---|---|
| 4 | 6. | 2. | | | | | | | | | | |
| 5 | 15.$^\square$ | 3.$^\square$ | 1. | | | | | | | | | |
| 6 | 44$^\triangle$ | 16.$^\square$ | 4. | 2. | | | | | | | | |
| 7 | 135$^\triangle$ | 36$^\triangle$ | 11.$^\square$ | 2.$^\square$ | 1. | | | | | | | |
| 8 | 528 | 128 | 28$^\triangle$ | 12 | 2. | 2. | | | | | | |
| 9 | 1354 | 275$^\triangle$ | 67$^\triangle$ | 20$^\triangle$ | 8 | 2. | 1. | | | | | |
| 10 | 4542 | 855$^\triangle$ | 175$^\triangle$ | 54 | 16$^\triangle$ | 8. | 2. | 2. | | | | |
| 11 | 14405 | 2457 | 477$^\triangle$ | 117$^\triangle$ | 36$^\triangle$ | 13$^\triangle$ | 5. | 2. | 1. | | | |
| 12 | 58976 | 14624 | 1369 | 924 | 83$^\triangle$ | 28$^\triangle$ | 11 | 4. | 2. | 2. | | |
| 13 | 167263 | 27376 | 3954 | 924 | 205$^\triangle$ | 61$^\triangle$ | 22$^\triangle$ | 9 | 4. | 2. | 1. | |
| 14 | 430080 | 192192 | 11878 | 2963 | 749 | 180 | 46 | 16$^\triangle$ | 7 | 4. | 2. | 2. |

The same approach can be taken to convert the problem of finding optimal weak $(n, d, w)$-DNA codes to a maximum clique problem on the graph $\Gamma_{n,d,w}^{GC}$, where the vertex set $V(\Gamma_{n,d,w}^{GC})$ is the set of all sequences $\sigma \in \Sigma^n$ with $GC(\sigma) = w$, and the edge set $E(\Gamma_{n,d,w}^{GC})$ contains all pairs $\{\sigma, \tau\}$ such that $d_H(\sigma, \tau) \geq d$.

We solve our maximum clique problem on $\Gamma_{n,d,w}^{GC,RC}$ and $\Gamma_{n,d,w}^{GC}$ using Cliquer, an implementation of Östergård's clique-finding algorithm by Niskanen and Östergård [35]. We found all optimal $(n, d, w)$-DNA codes for $(n, d, w) \in \{(5, 3, 2), (5, 4, 2), (6, 4, 3), (7, 6, 3)\}$ and all optimal weak $(6, 5, 3)$-DNA codes. The size of optimal DNA codes and optimal weak DNA codes for these parameter sets were not known previously [10]. The properties of the graphs $\Gamma_{n,d,w}^{GC,RC}$ and $\Gamma_{n,d,w}^{GC}$ for these parameter sets are given in Table I. The algorithm tends to perform faster on graphs of low *density* (a graph with $v$ vertices and $e$ edges has *density* $e / \binom{v}{2}$).

## V. COMPUTATIONAL RESULTS FROM STOCHASTIC LOCAL SEARCH

We use the lower bounds in the tables of Gaborit and King [10] and Tulpan [36] as benchmarks for the performance of our stochastic local search algorithm. Their table collects the best available lower bounds on $A_4^{GC,RC}(n, d, \lfloor n/2 \rfloor)$, the size of DNA codes having 50% GC-content. When $n \leq 14$, out of the 52 cases for which exact values of $A_4^{GC,RC}(n, d, \lfloor n/2 \rfloor)$ are not known, our stochastic search method yielded 20 record-breaking $(n, d, \lfloor n/2 \rfloor)$-DNA codes, showing its ability to improve upon results achieved by previous techniques, namely the stochastic search algorithms of Tulpan *et al.* [31], [30], the lexicographic search method of King [9], the coding theoretic method of Gaborit and King [10], and hybrid approaches combining the above methods [10]. The results obtained suggest that our method works well for the range $d \leq n \leq d + 6$, but other methods are better when $n \geq d + 7$.

Table II shows that state-of-art lower bounds for $A_4^{GC,RC}(n, d, \lfloor n/2 \rfloor)$ (entries followed by periods are exact values for $A_4^{GC,RC}(n, d, \lfloor n/2 \rfloor)$), with our results as follows:

1) numbers that are superscripted with $\triangle$ are new lower bounds obtained via the stochastic local search algorithm described in Section III;
2) numbers that are superscripted with $\square$ are exact values established via the maximum clique algorithm described in Section IV.

The DNA codes proving the lower bounds in Table II can be obtained from the first authors' website at

$\langle http://www1.spms.ntu.edu.sg/\sim ymchee/dnacodes.php \rangle$.

We point out that the lower bounds $A_4^{GC,RC}(12, 10, 6) \geq 4$ in the table of Gaborit and King [10] is in fact an equality. This follows from the values $A_4^{GC}(12, 10, 6) = 9$, and the halving bound in Lemma 1.

As a final remark, two of the new lower bounds for $A_4^{GC,RC}(n, d, w)$ obtained here yield new lower bounds for $A_4^{GC}(n, d, w)$ via Lemma 1:

$$A_4^{GC}(9, 5, 4) \geq 134,$$
$$A_4^{GC}(10, 4, 5) \geq 1710.$$

The previous best lower bounds for these were $A_4^{GC}(9,5,4) \geq 133$, obtained via simulated annealing [10], and $A_4^{GC}(10,4,5) \geq 1680$, obtained via the linear coding construction [10].

We note that algorithms similar or related to that proposed here have also been used with success on other coding problems (see, for example, [37], [38]).

## VI. CONCLUSION

We introduced a new stochastic search method for designing libraries of constant GC-content oligonucleotides satisfying combinatorial constraints necessary for reducing hybridization errors. With this new algorithm, we were able to improve on many of the benchmark lower bounds for DNA codes in [10]. The sizes of optimal DNA codes were also determined in several cases.

## ACKNOWLEDGMENT

The authors would like to thank their students X. Lu, Z. Shi, T. Wang, and Y. Wang, who found the DNA codes that prove $A_4^{GC,RC}(7,4,3) \geq 36$ and $A_4^{GC,RC}(8,5,4) \geq 28$, with an independent implementation of the algorithm described in this correspondence. They also thank the anonymous reviewers for their careful reading and insightful suggestions that helped improve the presentation of this correspondence greatly.